\documentclass[sigconf]{acmart} 
\AtBeginDocument{%
  \providecommand\BibTeX{{%
    \normalfont B\kern-0.5em{\scshape i\kern-0.25em b}\kern-0.8em\TeX}}}

\setcopyright{acmcopyright}
\copyrightyear{2024}
\acmYear{2024}
\setcopyright{rightsretained}
\acmConference[CHI '24]{Proceedings of the CHI Conference on Human Factors in Computing Systems}{May 11--16, 2024}{Honolulu, HI, USA}
\acmBooktitle{Proceedings of the CHI Conference on Human Factors in Computing Systems (CHI '24), May 11--16, 2024, Honolulu, HI, USA}
\acmDOI{10.1145/3613904.3642781}
\acmISBN{979-8-4007-0330-0/24/05}

\usepackage{tikz}  

\usepackage{tabularx}
\usepackage{multirow}
\usepackage{color}
\usepackage{listings}
\usepackage{color, colortbl}

\begin{document}

\title[Beyond Dark Patterns]{Beyond Dark Patterns: A Concept-Based Framework for Ethical Software Design}

\author{Evan Caragay}
\affiliation{%
  \institution{Massachusetts Institute of Technology}
  \streetaddress{32 Vassar St}
  \city{Cambridge}
  \state{Massachusetts}
  \country{USA}
  \postcode{02139}
}
\orcid{0009-0005-7286-5403}
\email{caragay@mit.edu}

\author{Katherine Xiong}
\affiliation{%
  \institution{Massachusetts Institute of Technology}
  \streetaddress{32 Vassar St}
  \city{Cambridge}
  \state{Massachusetts}
  \country{USA}
  \postcode{02139}
}
\orcid{0000-0003-0856-2123}
\email{kxiong22@mit.edu}

\author{Jonathan Zong}
\affiliation{%
  \institution{Massachusetts Institute of Technology}
  \streetaddress{32 Vassar St}
  \city{Cambridge}
  \state{Massachusetts}
  \country{USA}
  \postcode{02139}
}
\orcid{0000-0003-4811-4624}
\email{jzong@mit.edu}

\author{Daniel Jackson}
\affiliation{%
  \institution{Massachusetts Institute of Technology}
  \city{Cambridge}
  \state{Massachusetts}
  \country{USA}
  \postcode{02139}
}
\orcid{0000-0003-4864-078X}
\email{dnj@mit.edu}

\renewcommand{\shortauthors}{Caragay et al.}

\begin{abstract}
Current dark pattern research tells designers what \textit{not} to do, but how do they know what \textit{to} do? In contrast to prior approaches that focus on patterns to avoid and their underlying principles, we present a framework grounded in positive expected behavior against which deviations can be judged. To articulate this expected behavior, we use concepts---abstract units of functionality that compose applications. We define a design as dark when its concepts violate users' expectations, and benefit the application provider at the user's expense. Though user expectations can differ, users tend to develop common expectations as they encounter the same concepts across multiple applications, which we can record in a concept catalog as standard concepts. We evaluate our framework and concept catalog through three studies, illustrating their ability to describe existing dark patterns, evaluate nuanced designs, and document common application functionality. 

\end{abstract}



\maketitle

\section{Introduction}

When users experience deceptive design elements in applications, they may feel annoyed or frustrated---but they also feel resigned. There is a sense that deception is so rampant that users just have to live with it  \cite{maier_dark_2020}. To counter this, practitioners coined the term ``dark pattern'' \cite{brignull_dark_2010} to label the common tactics and idioms that are employed to deceive users, cataloging them and pointing to examples, in the hope that exposing them would lead responsible designers to shun them.

Recognizing the importance of this problem, researchers have investigated the incidence of dark patterns \cite{soe_circumvention_2020, zagal_dark_2013, gunawan_comparative_2021, kowalczyk_understanding_2023, bosch_tales_2016, hidaka_linguistic_2023, mathur_dark_2019, mildner_about_2023}; explored how users react to them \cite{di_geronimo_ui_2020, nouwens_dark_2020, maier_dark_2020, utz_informed_2019, luguri_shining_2021, chordia_deceptive_2023} and how designers choose to incorporate them (or not) \cite{sanchez_chamorro_ethical_2023, chivukula_dark_2018}; and more broadly have developed theories seeking to articulate exactly what makes a design ``dark'' \cite{mathur_what_2021, ahuja_conceptualizations_2022}. In so doing, a gap has opened between this rich body of research and the working practices of designers \cite{gray_dark_2018}.

This is in part due to a cultural divide, with ethical discussions more prevalent in academic settings, and in part because ethical guidelines are not easily translated into practice, especially when they require subtle judgments about borderline cases.

Less recognized, but perhaps equally important, is the fact that practitioners spend less time making fresh design decisions than academics often assume. Much of their effort involves adoption (and  adaptation) of large pieces of functionality and interaction design. In this respect, the pattern-driven approach that says ``don't do things like this'' may be more practical than theory-driven approaches that offer more general criteria.

Pattern catalogs and theories thus have complementary benefits. Catalogs are readily actionable: their contents are easily evaluated, and they can usually be matched to given circumstances and applied straightforwardly. Theories, despite being less immediately applicable, provide the underlying principles and ethical justification for consensus viewpoints, and a language for exploring edge cases when consensus is harder to come by.

The existing catalogs of dark patterns, however, have limitations. First, and most obviously, they tell a designer what \textit{not to do}, when a designer is usually seeking advice on what \textit{to do}. This not only makes a catalog less useful but also means that the larger community of designers is missing an opportunity to publicize and inculcate good practices.

Second, although the match between a pattern and an exact instantiation may be  clear, it can be harder to evaluate variant designs that share only some but not all features of the pattern, or that present them in subtly different ways. It is unclear, for example, whether the addition of ``suggested items'' to a user's shopping cart is a violation of the \textit{Sneak into Basket} dark pattern \cite{brignull_deceptive_2023}.

Third, a catalog of negative patterns is necessarily incomplete. Ensuring the \textit{absence} of a particular set of dark patterns from a design cannot be sufficient for establishing that the design is acceptable. Herein lies the challenge of regulating against dark patterns. A well-intentioned company that has avoided known dark patterns in its products is surely entitled to some confidence that it will not be sued for the inclusion of design elements that might come to be regarded as dark but were never explicitly specified. This problem is exacerbated by the broad nature of some regulatory efforts (for example, the suggestion in a recent EU white paper \cite{eu_dark_2022} that all forms of emotional steering are culpable).

In this paper, we outline a new approach that combines aspects of both patterns and theory. The key idea is to define acceptable designs as instantiations of positive patterns, deviations from which are treated as ``dark.'' Our patterns are based on \textit{software concepts }\cite{jackson_essence_2021} which were introduced as a way to modularize application functionality.

A concept is a unit of functionality that offers a coherent and separable service to the user, fulfilling a clear purpose. An online store's concepts might include \texttt{ShoppingCart}, \texttt{Order}, \texttt{Catalog}, etc., as well as less domain-specific concepts such as \texttt{UserAuthentication}, \texttt{Notification}, \texttt{Recommendation}, etc. We propose the creation of \textit{concept catalogs} that document standard versions of concepts and their variants.

The contributions of this work are:
\begin{enumerate}
\item \textbf{A framework for evaluating designs}. Regulators, designers (and critics more generally) can use our approach to evaluate  designs. Roughly, a design is dark when the user's expected concept does not match the underlying application concept. Many prior approaches likewise define darkness in terms of deviation from user expectations, but do not indicate how those expectations are obtained. We show how \textit{concepts} can play this role.
\item \textbf{A language for consensus standards}. Many industry groups have benefited from defining  standards that capture best practices; notable examples include the PCI standard for ensuring payment security \cite{noauthor_standards_2023} and Web accessibility standards \cite{lawton_henry_wcag_2023}. \textit{Concept catalogs} can provide a structure and language for articulating conventional behaviors and interfaces for common functional units such as shopping carts.
\item \textbf{A design method}. As a method (for designers and developers), our approach suggests a separation of design work into two aspects: the collaborative creation of catalogs to describe existing concepts, and the instantiation of concepts from a catalog in the context of a particular design. This is the programme Alexander proposed in the context of architecture \cite{alexander_pattern_1977}, and that was realized in software (albeit at the code level) in the community that grew out of the design patterns book \cite{gamma_design_1994}.
\end{enumerate}

Our paper begins by reviewing existing taxonomies and theories of dark patterns, pointing to differences and synergies between them and our approach (Section \ref{02-relatedwork}). We explain our approach, starting with the idea of software concepts (Section \ref{03-background}), defining criteria for darkness based on deviation from conventional concepts (Sections \ref{04-definitions}, \ref{04-deviation}), and then presenting the idea of catalogs of concepts (Section \ref{05-catalog}).

We evaluated our approach in three studies. First, we checked that it can account for uncontroversial negative examples, by applying it to the latest version of Brignull's catalog (Section \ref{06-brignullcoverage}). Second, we analyze three shopping cart variants to test its ability to make nuanced distinctions (Section \ref{06-shopping-case-study}). Third, to check that a catalog concept can accommodate the real variety of instances, we construct a sample concept and make sure it accommodates all the variants found in a repertoire of popular web applications (Section \ref{06-popularwebsites}).

Finally, we discuss the merits of our approach (Section \ref{07-merits}); some challenges to it (Section \ref{07-limitations}); and conclude with some ideas about next steps (Section \ref{07-nextsteps}).

\section{Related Work}
\label{02-relatedwork}

\textbf{The origin of dark patterns}. Harry Brignull, who introduced the term dark patterns\footnote{Some have proposed replacing this term with alternatives such as “deceptive design” or “manipulative design”. We chose to not use these, as our framework explicitly does not rely on deception or manipulation to evaluate dark patterns, but we recognize the need to consider more inclusive terminology. We acknowledge that the ACM Diversity and Inclusion Council's list of terms to be avoided now includes the term “dark patterns” (https://www.acm.org/diversity-inclusion/words-matter).} in 2010, defines them as “tricks used in websites and apps that make you do things that you didn’t mean to” \cite{brignull_dark_2010, brignull_deceptive_2023}.  Since then, much work has appeared, identifying and categorizing dark patterns and the strategies they  employ, along with taxonomies, some seeking to be applicable across all designs and some tailored to specific domains \cite{brignull_dark_2010, bosch_tales_2016, gray_uxp2_2022, mathur_dark_2019, zagal_dark_2013, hidaka_linguistic_2023, gunawan_comparative_2021}.

\textbf{Positive vs. negative patterns}. Our approach is most similar to these approaches, to the extent that they are inspired by Christopher Alexander's influential idea of patterns \cite{alexander_pattern_1977}: that design knowledge can be organized around archetypal design forms. But whereas dark patterns focus on negative archetypes (what one should not do), our concepts focus on positive archetypes (what one should do), in common with Alexander's patterns. This approach is more easily aligned with standard design practice, since by adopting a standard concept, a designer can be reassured that their work is acceptable, without having to check the design against a potentially boundless (and changing) collection of negative patterns.

\textbf{Defining darkness}. Sensing the ad hoc nature of dark pattern taxonomies, Mathur et al. pointed to a missing definition of ``what, exactly, makes a user interface design a dark pattern'' \cite{mathur_what_2021}. They note the consequences of this gap, including limited dialogue across disciplines, and a disconnect between academics and regulators. Mathur et al.’s concerns reflect a more general shift in recent scholarship, away from descriptive taxonomies and towards understanding the specific underlying mechanisms that dark patterns use.

\textbf{Identifying underlying strategies}. Researchers have tried to identify common strategies that underlie dark patterns and might provide a unifying explanation. Building on Thaler and Sunstein's notion of ``choice architecture'' \cite{thaler_nudge_2008}, Mathur et al. defined dark patterns as modifications of the standard choice architecture that disadvantage the user by manipulating the available information or modifying the available choices \cite{mathur_what_2021}. With a similar focus on the user’s experience of the design, Gray et al. identified two dimensions that impact a user’s “felt experience of the design”: the “user’s interpretation of the system [...], including relevant affordances” and “the expected outcome that the user anticipates as a result of their previous or present interaction with the system” \cite{gray_dark_2018}. They helpfully identified five archetypal strategies that characterize dark patterns (nagging, obstruction, sneaking, interface interference and forced action).

\textbf{Avoiding a reliance on underlying strategies}. Our approach does not require identifying the underlying psychological strategies used to influence a user's choices. A design is dark by virtue of deviating from standard expectations in a manner that is contrary to the user's interests. Thus, we can account for designs that many users find acceptable, despite the fact that they employ nudges at the user's expense. For example, online retail sites routinely try to sell extended warranties to consumers, despite the fact that such warranties are usually overpriced. Most consumers are not too troubled by these tactics, because they expect them: a nudge to buy an extended warranty is no darker to them than the placement of overpriced chocolates at supermarket registers.

\textbf{Defining the default choice architecture}. Mathur et al.’s observation that dark patterns arise from modifications of a default choice architecture raises the question of what that default choice architecture is. It seems unlikely that one could find a general criterion to answer this question; requiring all choices to be presented uniformly, for example, would clearly not work, since in all designs some choices are more commonly made than others, and for this reason alone are typically emphasized.

Our approach resolves this conundrum by introducing concepts to capture  conventional practice and represent the common expectations of users.  This is in line with Gray et al.'s two dimensions of a user’s felt experience of a design (expected outcome and current interpretation / available affordances). But whereas these dimensions are secondary for Gray, being observations of the discomfort that users suffer from as a consequence of the underlying strategies that they identify, in our approach the deviation from expectation is primary, independent of whatever strategy might have produced it.

\textbf{Combining procedural and substantive ethics}. Mathur et al. tie choice architecture modifications to how those modifications impact normative values. Our approach likewise offers a criterion that combines a more simple and objective criterion (in our case deviation from the standard concept) with a more subjective one (namely whether the deviation is contrary to the user's interests). In this way, our work  combines two kinds of ethical criteria: \textit{procedural}, which are cast in terms of standardized and repeatable steps, and \textit{substantive} which are concerned with more subjective judgments that depend on context \cite{zong_bartleby_2022}.

\textbf{Judging nudges, good and bad}. Other research aims to distinguish ethical from unethical “nudges.” Meske and Amojo define criteria: a nudge should be transparent, resistible, and non-controlling \cite{meske_ethical_2020}. Sánchez Chamorro et al. focus on salience and force as the underlying mechanisms; what distinguishes whether the influence of a design as dark or not is whether the influence is apparent vs. hidden, or strong vs. weak \cite{sanchez_chamorro_ethical_2023}. These distinctions are insightful and  useful, but may not provide sufficient grounds for deeming a design to be dark or not. In our approach, the focus is not on whether a nudge is present, or what form it takes, but whether the behavior that the user is being nudged to perform is conventional (that is, in line with the concept) and in the user's interests. This allows us, in particular, to be more permissive in allowing nudges to enforce standard and desirable behaviors, such as pushing users to backup their data or change passwords that have been compromised.

\textbf{User expectations in existing work}. User expectations have also been explored elsewhere in the literature. Sánchez Chamorro et al. discuss how UX practitioners already reflect on user expectations when designing, drawing upon “the design strategies and mechanisms that users are familiar with” \cite{sanchez_chamorro_ethical_2023}. Their observation that “Designers adjust their designs to standards that the user might expect, and the user, in turn, gets accustomed to the designs they interact with” aligns closely with concept design and our framework. Other research highlights the non-homogeneity of user expectations and the impact of context \cite{hallinan_unexpected_2020}.

\section{Dark Patterns as Deviation from Standard Concepts}
\label{sec:theorizing}

\subsection{Concepts}
\label{03-background}
In this section we introduce \textit{software concepts}, which form the foundation of our framework. In \textit{The Essence of Software} \cite{jackson_essence_2021}, Jackson introduced concepts as a way to structure the functionality of a software design independently of its implementation architecture.

Concepts are abstract, self-contained units of functionality, each of which fulfills a specific purpose. A social media app might be described in terms of concepts such \texttt{Post}, \texttt{Comment}, \texttt{Upvote}, \texttt{Friend}, etc. Concepts are generic and can be applied in multiple contexts: thus the \texttt{Upvote} concept, for example, can be applied both in a social media app for upvoting posts, and in a question-and-answer forum for upvoting answers.

A concept can be modeled as a state machine comprising a (possibly infinite and structured) set of states, and some actions that read and write the state. For example, the \texttt{User} concept that governs user authentication in most apps has a state comprising names and passwords for each user, and actions (1) to register a new user and (2) to authenticate with username and password (which fails if the wrong combination is entered).

Concepts are composed by \textit{synchronization}, in which the actions of different concepts are tied together. Using synchronization rather than conventional procedure call ensures that concepts remain decoupled from one another. For example, user authentication might be applied to posting by composing the \texttt{User} concept with the \texttt{Post} concept, so that the action of \texttt{Post} that creates a new post is synchronized with the authentication action of \texttt{User}. This will ensure that a new post cannot be created without authentication, and (by passing the user identity in the synchronization) will also ensure that the author of the post will be the authenticated user.

Concepts are experienced in the user interface through its views and controls. To describe this, we associate with each concept in an app a \textit{mapping} that defines how the actions of the concept are mapped to user interface controls, and how the state of the concept is mapped to views. This will be important because in dark patterns problems can arise separately from the underlying functionality (the concept per se) and the design of the user interface (how the concept is mapped).

This notion of mapping is a realization in the context of concept design of the idea that a user interface ``projects'' an image of the designer's conceptual model, and that problems arise when that projection is not faithful and accurate\cite{norman_design_1998}. Concept design brings two new aspects to this approach. First, the granularity of concepts allows these projections to be evaluated on a concept-by-concept basis. Second, both the underlying concepts and their mappings are standardized, so that mappings do not need to be considered afresh for each application. There are completely standard ways, for example, to represent the actions and state of the \texttt{Upvote} concept (with thumbs-up and thumbs-down widgets, for example, and visible vote counts).

In summary, a concept description includes:
\vspace{-\topsep}
\begin{itemize}
    \item \textit{State}. A concept's state keeps track of values that may potentially change with user interaction. For example, a \texttt{Catalog} concept for an online store maintains a collection of items that are available for sale, along with their title, price and other characteristics. A separate \texttt{Inventory} concept might track the availability of particular items; its state would count the number of each item available (and perhaps characteristics of item instances, such as serial numbers).
   
    \item \textit{Actions}. Actions characterize the behavior of a concept; state is modifiable only through actions. The actions of \texttt{Catalog} would include those for listing and delisting items and changing their characteristics; the actions for \texttt{Inventory} would include actions for updating the inventory as items are acquired and sold. Actions may have preconditions that limit their occurrence: in a \texttt{Shopping Cart}, a user can only remove an item from their cart if the item is already in the cart. In mixed-initiative systems, there may also be constraints on who can initiate an action (for example, that items cannot be added to a cart except by users).

    \item \textit{Synchronizations}. A synchronization ties actions from different concepts together. For example, the action of the \texttt{Order} concept that is executed when an order is fulfilled would be synchronized with the action in the \texttt{Inventory} concept that causes the inventory to be appropriately reduced.
    
    \item \textit{Mapping}. States and actions are mapped to their concrete representation in a user interface. A good mapping results in a user interface that accurately and transparently communicates the concept’s functionality. The state of a concept need not be fully visible to all users; for the \texttt{Inventory} concept, for example, the user interface of the online store will have to indicate whether or not a particular item is in stock or out of stock, but the actual number of instances available may only be visible to company employees.
\end{itemize}
A full example of a concept definition (for \texttt{Shopping Cart}) is shown later in \autoref{concept-ex}.

\subsection{Observed, Implemented, and Expected Concepts}
\label{04-definitions}

To articulate users' expectations, we identify three variations of a concept: the \textit{observed concept}, the \textit{implemented concept}, and the \textit{expected concept}. By comparing these, we can identify whether a design aligns with user expectations. If a design violates those expectations, analyzing differences in these concepts lends further clarity to users' perceptions of how ethical the design is, and can suggest changes to re-align expectations.

The \textbf{observed concept} is the concept that the user observes when they interact with the application. It is formed from the state and actions a user sees in the user interface, and represents their interpretation of how the application works, regardless of the intentions of the designer or programmer.

The \textbf{implemented concept} is the concept that is implemented in the actual functionality of an application, establishing which actions are possible to take and how those actions impact the state. It does not depend on the user at all but is inherent in the application itself. When a concept is mapped well to a user interface, the observed concept will closely match the implemented concept; all the actions will be visible to the user and the user will have a clear and accurate understanding of the state. However, when a user interface does not adequately communicate the implemented concept, the observed concept will differ due to gaps in the user's understanding. 

For example, a subscription service may have an implemented concept that enables both subscribing and unsubscribing.
But if the mapping makes unsubscribing difficult, the user’s \textit{observed} concept may not include an unsubscribe action.
In cases where an application contains multiple concepts, a UI might fail not only to show certain states and actions, but also to communicate the presence of entire concepts.

The \textbf{expected concept} is the concept that a user expects to be interacting with, based on their experiences with that concept in the past. Common concepts, like \texttt{Account} or \texttt{Shopping Cart}, appear on enough websites that users have well-formed expectations for how they work. The expected concept includes both the behavior the user expects at the moment of interaction, along with the consequences of that interaction in the future.

These distinctions also apply to compositions of concepts, both in how concepts are composed (that is, which synchronizations are present) and even which concepts are included. For example, \texttt{EmailMessage} and \texttt{Label} are frequently used in conjunction with each other, so users have developed expectations for their synchronizations: that when an email is sent, the ``sent'' label is added to it, eg.

\subsection{Defining Deviation}
\label{04-deviation}

Building on these conceptual notions, we can now present our definition of what comprises darkness. A design is dark for a given user when:
\begin{enumerate}
    \item The observed or implemented concept is different from the user's expected concept;
    \item The difference benefits the designer or owner of the system to the detriment of the user.
\end{enumerate}

Consider, for example, a shopping cart in which items are added spontaneously without the approval or even awareness of the user. Such behavior satisfies both conditions for darkness. The behavior is both (1) unexpected, since the user's expected concept of a shopping cart does not involve items being added without their participation, and (2) to the detriment of the user, since they risk purchasing items they don't want. One might counter that the company is providing the user with helpful purchase suggestions. Such functionality, however, would better be offered as part of a \texttt{Recommendation} concept, which would have the desired benefit while both being familiar to the user and not causing the user any risk of accidental purchase of unwanted items. 

On the other hand, consider how many shopping carts only show shipping costs at the end. This meets the second criterion of acting to the detriment of the user, since it would undoubtedly be more helpful for the user to know shipping costs upfront before they go to the trouble of selecting items. In this case, however, the first criterion is not met, since this non-optimal behavior is what users typically expect.

Now suppose the shopping cart were designed so that the shipping cost were included and updated incrementally as more items are added. This is not typical: it's complicated to implement, shipping costs require knowledge of the location to which the order will be sent, and different shipping options might be available. So this design would meet the first criterion, being a deviation from expectation. But it would fail the second criterion, since the deviation is not to the detriment of the user. On the contrary, the user benefits from an earlier estimate of shipping costs.

Our two criteria have a different flavor from one another, and represent \textit{procedural} and \textit{substantive} aspects of ethics \cite{zong_bartleby_2022}.
Arguments that interfaces are ethical if they don't include certain kinds of design patterns, or if designers follow certain steps to enable individual choices, are procedural; arguments about the nature of harm to the user are substantive.  We believe that both perspectives are essential to a complete understanding of dark patterns.

\subsubsection{Procedural Considerations}

We now consider the procedural aspects of each kind of deviation.

\textbf{Observed concept violates expected concept}: When the observed concept breaks a user's expectations of an existing concept’s behavior, the user interface is communicating functionality that violates the user's preconceptions of how the application will behave. This violation could be due to an action not being available to the user, a state not representing the expected state, or other changes to the concept's functionality.

\textit{Example: }Brignull's ``hard to cancel'' dark pattern occurs when ``the user finds it easy to sign up or subscribe, but when they want to cancel they find it very hard'' \cite{brignull_deceptive_2023}. In this case, \texttt{Subscription} is the relevant concept. The user's expected concept is that \texttt{Subscription} has both a subscribe and unsubscribe action. However, when the user interacts with the application, their observed concept is that unsubscribe is not available. Even if the implemented concept includes an unsubscribe action (such that it is technically available), if the user finds it very difficult to access, their observed concept will not include the action. Therefore, the observed concept violates the expected concept.

\textbf{Implemented concept violates expected concept}:  In some cases, the observed concept, as projected by the user interface, appears to conform to the concept the user expects, but the underlying behavior is not as expected. In this case, the expected concept matches the observed concept, but not the implemented concept. This mismatch may not become clear to the user immediately (for example, placing an online order but the item never arrives), may only be clear to some users (for example, a bug that causes certain items not to be added correctly to a shopping cart), or remain hidden indefinitely (for example, a recommendation system claiming to rank content by relevance, but actually showing sponsored content). 

\textit{Example: }Brignull’s examples of fake scarcity, fake social proof, and fake urgency \cite{brignull_deceptive_2023}. In fake scarcity, the expected concept is that the state of the concept (ex: an \texttt{Inventory}'s count of items) is accurately displayed. But the implemented concept does not behave as expected: instead of displaying an accurate count of items in the \texttt{Inventory}, the UI displays an arbitrary number. In this case, the observed concept fulfills the expected concept, but the implemented concept violates the expected concept. 

\subsubsection{Substantive Considerations}
\label{sec:substantive}

Deviating from an expected concept isn't inherently unethical---it matters what outcome is produced by that deviation, and whether users are harmed by that outcome.

While both procedural and substantive perspectives are essential to a complete understanding, prior work in dark patterns has typically been more procedural. While Mathur et al.  call for more normative approaches to dark patterns (which would be considered substantive), their recommended normative considerations tend to focus on a limited set of liberal values (for example, autonomy and fairness in markets) \cite{mathur_what_2021}. As data ethics scholars have noted, there are other dimensions of justice that are not included within a rubric of rights, opportunities, and material resources \cite{hoffmann_where_2019}.
Instead of universalizable normative concerns, we recommend a more situated, context-specific approach centered in users' subjective expectations and values.

Space does not permit a full analysis of these substantive issues. They include recognition that users have subjective values; that their values may conflict with the designer's values; that the values of different users may not be simultaneously satisfiable; and that individual autonomy may conflict with the common good.

\section{Concept Catalogs: Articulating Standard Expectations}
\label{05-catalog}
Concepts allow us to identify common expectations of functionality. But concepts are often implicit and not documented explicitly. To enable designers and policymakers to articulate concepts that can be referenced, implemented, and critiqued, there is a need to define standard concepts that reflect commonplace expectations.

In this section, we explain the idea of \textit{concept catalogs}, a tool for formally defining, agreeing upon, and communicating standard expectations for concepts, and we discuss how designers can map a standard concept to a user interface. We show how to create a concept catalog entry via an example.

Building the catalog from concepts with a focus on functionality allows for more precise design analysis, in comparison to existing tools such as platform design guidelines (e.g. Google Material Design \cite{noauthor_material_2023} and Apple Human Interface Guidelines \cite{noauthor_human_2023}) or design pattern catalogs (e.g. ui-patterns.com \cite{uipatternscom_2023}, GoodUX \cite{kim_goodux_2023}, UXarchive \cite{noauthor_uxarchive_2023}). These tools tend (for our purposes) to be too specific, being limited to a particular platform or UI framework, or too broad, focusing on the generality of UI widgets without considering the functionality behind them.
 
Even without consensus on what the standard expectations should be for a certain concept, a concept catalog entry allows designers to explicitly articulate a point of reference that they are designing toward.
The systematic and reusable nature of concepts allows the defined expectations to be both precise and applicable across domains. Standard concepts can provide the language to ground critiques during design discussions and give designers a baseline to work from. They can also help companies meet a consistent set of expectations across their products, and help regulators communicate and enforce expectations that are particularly critical to protecting users.

\subsection{Components of a Concept Catalog Entry}
\label{sec:concept-catalog-entry}
A concept catalog entry contains the concept definition and the concept mapping. Together, they specify the standard core functionality of a concept and its standard mapping to a user interface.

\textbf{Concept Definition.} The concept definition should enumerate the concept's state and actions. It should also define expected synchronizations---or, when an action in one concept results in an action being called in a different concept. For mixed-initiative systems, the definition should specify who is allowed to initiate each action.

Different implementations of the same concept may include extensions of the concept's core functionality. A standard concept is meant to capture the minimal functionality that is essential to forming users' expectations.
Almost all applications will implement extensions beyond this minimal definition, but these extended implementations must not conflict with the expectations established in the minimal standard concept. Scoping standard concepts to only include core functionality enables designers to use the catalog while maintaining the flexibility to innovate and extend their applications. The catalog will also generally outline common variants, but these are treated as optional.

The concept definition can be articulated in a formal concept language (as in \autoref{concept-ex}) or informally. While defining a standard concept in plain language is often more useful to designers using the entry due to readability, defining a formal concept requires additional precision, which can be helpful for improving the quality and completeness of the concept. 

\textbf{Concept Mapping.} For each action and state within the concept definition, the catalog entry can  specify standards for mapping the concept to a user interface. A mapping exposes a concept's actions and state as user interface components, like buttons and text labels.
Just as users form expectations about functionality, they also expect mappings to follow familiar patterns. If a user encounters an unexpected mapping, their observed concept may be inaccurate or reflect a different concept entirely. To prevent this, catalog entries can specify standard mappings for a given concept that designers can use.

While there are general usability principles that apply to mapping, which we discuss in \ref{mapping_principles}, there may be mapping requirements that are specific to a concept or a context of use. Mapping standards only need to describe the minimum requirements for the UI to convey the concept's functionality to a user---they do not include implementation details or stylistic guidance.

\subsection{Shopping Cart: An Example Catalog Entry}

In this section, we define an example \texttt{Shopping Cart} standard concept and provide example mapping standards. We show the essential parts of the formal definition in \autoref{concept-ex}. A full version can be found in \autoref{appendix}.

\textbf{Concept Definition}
The concept's states, actions, and synchronizations are written to be reusable (applicable to any shopping cart) and minimal (only the required, core parts of the concept are defined).

\textit{States.}
The \texttt{Shopping Cart} concept has state linking each cart to an owner and a set of items. Each item has a quantity and price, from which the subtotal is derived.

\textit{Actions.}
The \texttt{Shopping Cart} concept defines actions required to update its state, along with whether the user or shop can initiate each action. These actions include initializing the cart, adding and removing items, updating item quantities and prices, and checking out.

\textit{Synchronizations.}
The \texttt{Shopping Cart} is synchronized with \texttt{Catalog} (to ensure matching prices and accurate stock counts) and \texttt{Order} (to create an order when a shopping cart is checked out). Defining standard synchronizations help the implementation match users' expectations of functionality that spans multiple concepts.

\textit{Concept Mapping.}
\label{05-concept-mapping-example}
The mapping standards for \texttt{subtotal} and \texttt{quantityChange} are examples of state and action mapping standards. These standards provide positive guidance, describing how the UI should behave rather than listing UI elements to avoid. Additionally, they focus on functionality instead of prescribing implementation details such as components, style, or copy-writing.

\begin{figure*}[t]
 \begin{minipage}[b]{0.36\textwidth}
     \centering
     \includegraphics[width=1\linewidth]{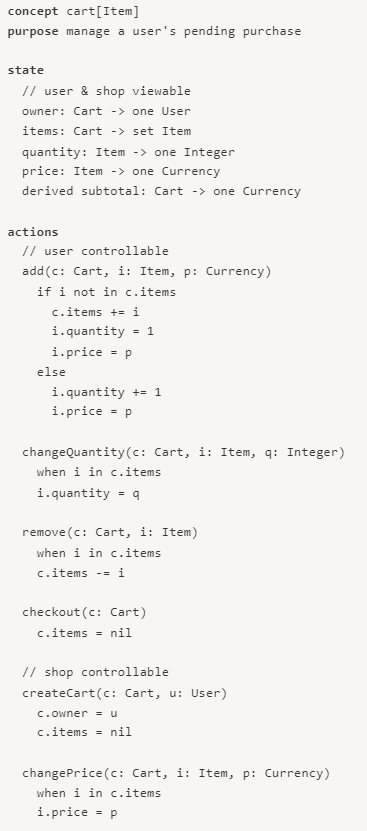} \newline
     (A)
    \end{minipage}\hfill
    \begin{minipage}[b]{0.61\textwidth}
     \centering
     \includegraphics[width=1\linewidth]{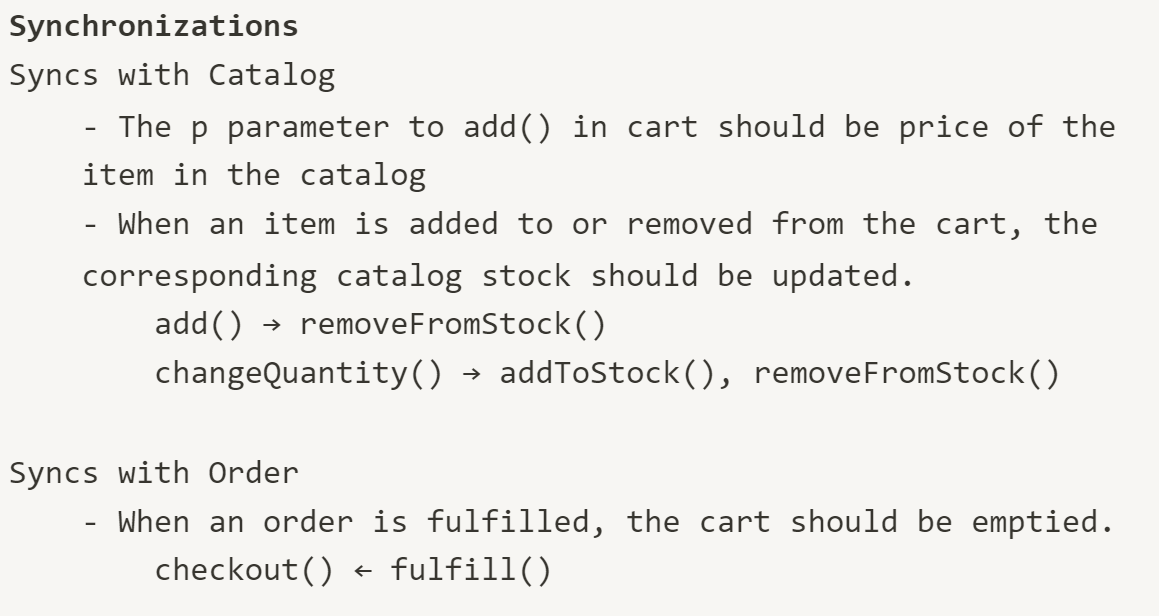} \newline
     (B)

    \centering
     \includegraphics[width=1\linewidth]{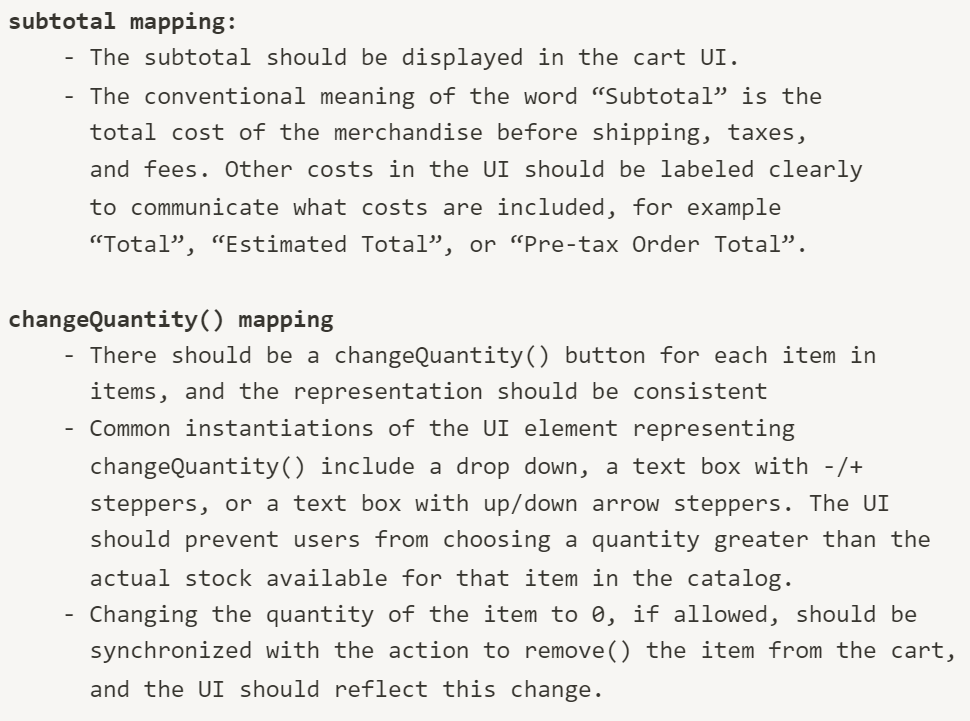} \newline
     (C)
    \end{minipage}

    \caption{Shopping cart concept catalog entry that designers can can compare to check if their design fulfills the requirements set by the standard concept. (A) Standard, formal concept definition, articulating the purpose, state, and actions. (B) The required synchronizations for the shopping cart, which helps designers meet user expectations of functionality between concepts, in addition to within individual concepts. (C) Example standards for how to map the state (ex: subtotal) and actions (ex: quantity change) to the user interface.}
    \label{concept-ex}
    
    \Description{All components of the figure are described in the text of section 5.2.2.}
\end{figure*}

\subsection{Refining the definition of dark pattern}

The definition of darkness in the presence of standardized concepts needs some refining. We shall say that a dark pattern occurs when:

\begin{enumerate}
    \item The observed or implemented concept does not include the standard concept's functionality
    \item The difference from the standard concept benefits the designer or owner of the system to the user's detriment
\end{enumerate}

As with our previous definition using expected concept, using standard concepts allows us to think in terms of where an application might diverge from standard expectations of functionality. 
While different standards may emerge for the same concept, organizations can develop catalogs to articulate the standards they intend to follow. Designers and researchers can use the standard concept definition appropriate to their context. Regulators and organizations can also compare and evaluate different ways of defining standard concepts, advocating for the adoption of definitions that have been thoroughly reviewed.

\subsection{Mapping Principles}
\label{mapping_principles}

Concept catalog entries include concept-specific standard mapping guidance, but there are also general mapping principles that apply to all concepts.
Here, we discuss a few relevant  principles (based on well known usability heuristics \cite{norman_design_1998, nielson_10_2020, tognazzini_first_2014}) that help designers clearly communicate a concept's state and actions.

\subsubsection{Transparency.} The UI should accurately and fully convey all parts of the concept that are available to the user.
    \begin{itemize}
        \item \textbf{Correspondence}: There should be a 1:1 mapping between the standard concept's user-viewable states and user-controllable actions, and visible user interface elements. 
        In other words, each state and action should correspond to visible and accessible UI elements, and there should be no elements that imply states or actions that are not part of the concept.
        \item \textbf{Faithfulness}: The UI should truthfully convey the value of each state component, and accurately convey what actions are triggered by each interactive element.
    \end{itemize}
\subsubsection{Uniformity.} The mapping should represent the underlying concept in a uniform manner:
    \begin{itemize}
        \item \textbf{Consistency}: When the same state or action is represented in multiple places, those representations should all look and behave the same.
        \item \textbf{Symmetry}: Actions that represent alternative options for a choice (e.g., accept and decline) should be given equal prominence in the interface.
        \item \textbf{Conventions}: Some states and actions have conventional representations that are well established. For example, an ``x'' typically represents cancel or exit. If these representations are used, they should represent the expected state or action. (This is arguably another kind of consistency, albeit across applications.)
    \end{itemize}

\section{Evaluation}

In this section, we evaluate our two contributions---the framework and the concept catalog---in three small studies. Each study was designed to address a particular research question that concerned us:

\textbf{RQ1:  Can previously identified dark patterns be accurately described by the introduced framework?} The introduced framework takes a different approach to defining dark patterns than past approaches; if it cannot describe existing patterns, then the applicability of the framework would be limited. For example, we were concerned that there might be established dark patterns that cannot be explained as concept deviations, maybe because they don't involve a deviation from user expectations, or because no concept from which they deviate can be readily identified.

To address this question, we re-described each dark pattern in Brignull's Deceptive Patterns taxonomy \cite{brignull_deceptive_2023}. We selected this taxonomy because it was recently updated in April 2023 to include major dark patterns documented in the literature, including all patterns identified by Gray et al. \cite{gray_dark_2018}, Mathur et al. \cite{mathur_what_2021}, Bosch et al. \cite{bosch_tales_2016}, and by Brignull's original list \cite{brignull_dark_2010}. 

\textbf{RQ2: Can the introduced framework evaluate nuances in potentially dark design that current taxonomies fail to capture?} We had an intuition that our approach might not only cover existing dark pattern identifications, but might sometimes allow more nuanced distinctions. Early on in our research, we constructed a list from the literature of 150 known examples of dark patterns, and we took these as challenges to our evolving criteria as we developed them. We had identified three particular websites that seemed similar but which we were initially unable to distinguish cleanly. We therefore chose to revisit these to check that our framework was indeed capable of resolving these cases. This demonstration is existential and not universal; success here only means that some additional nuances can be captured, not all. 

We explored the framework's evaluative and generative power through a case study comparing three applications that are superficially similar. We examined user reviews of the sites for evidence that users do not perceive these sites to be equally unethical. We then evaluated the sites' designs using existing taxonomies and our introduced framework, identifying where our introduced framework contributes new insights and where it fails to capture nuances. Further, we evaluated the framework's generative power by showing how designers can use the specific divergence from the expected concept to identify where they need to realign expectations.

\textbf{RQ3: Do websites share common functionality patterns and can a concept catalog entry succinctly capture these common patterns?} If common functionality across websites is not as consistent as concept theory proposes, then defining standard concepts would not be a feasible approach to combating dark patterns. Further, if the functionality of concepts was too complex or nuanced to capture, the applicability of the catalog to real world design would be limited.

First, one researcher constructed a catalog entry of the standard shopping cart concept. They surveyed (directly and through API documentation) the eCommerce platforms with the most global market share (WooCommerce, Squarespace, and Shopify) \cite{noauthor_13_2021}, along with top eCommerce sites such as Amazon and Etsy. In addition to examining top sites, the researcher used search terms similar to “eCommerce Shopping Cart” and “[specific feature] best practices” to identify blogs providing design advice, along with a variety of search terms on Google Scholar, such as “Shopping Cart ecommerce” and “Shopping Cart features,” to identify relevant academic literature. Through these methods, the researcher defined identify common functionality to include within the base concept.

Second, a \textit{different} researcher conducted a review of top e-comm- \\ erce sites to identify whether the \texttt{Shopping Cart} catalog entry accurately describes functionality across sites. They manually examined the top 20 e-commerce websites in the USA as of June 2023, ranked by web traffic and engagement on Similarweb \cite{noauthor_top_2023}. Similarweb is an industry-standard web traffic analytics platform \cite{jansen_measuring_2022} that reportedly has data on over 100 million websites \cite{noauthor_similarweb_2021}. They excluded e-commerce websites that did not contain the shopping cart concept (e.g. Rakuten), as well as websites whose primary language was not English. Table \ref{table:websitetable} shows the top 20 remaining websites. 

Of course affirmative answers to these research questions are \textit{necessary} for our approach to be successful but not \textit{sufficient}. We discuss in the limitations section (\ref{sec:study-limitations}) particular respects in which these studies are incomplete.

\subsection{RQ1: Coverage of Existing Dark Pattern Taxonomies}
\label{06-brignullcoverage}
\autoref{table:brignulltable} identifies the concepts most commonly associated with each of Brignull’s deceptive designs and briefly explains how our framework characterizes the dark patterns in terms of a violation of the expected concept, answering RQ1 in the affirmative.

For instance, some dark patterns are caused by explicit misrepresentation when the underlying concept is mapped to a user interface. These cases include comparison prevention, disguised ads, fake scarcity, fake social proof, fake urgency, hard to cancel, and hidden costs. The implemented concept will always be different from the expected concept, as user expectations are formed by a deceptive user interface.

In addition to covering each dark pattern in Brignull's taxonomy, our framework can also offer additional descriptive power. Where Brignull's list gives equal weight to each entry in the taxonomy, in some cases using concept analysis lets us differentiate between various reasons a pattern could be considered dark.

While the explicit misrepresentation cases are egregiously dark, other dark patterns depend on subtler judgments of user expectation; in these cases, concept theory is able to describe nuances that are not captured by the taxonomy. 
Many dark pattern definitions implicitly assume that user expectations are independent of context. For example, the nagging pattern reflects the expectation that a disable action should always be available whenever there are repeated requests to the user. But such an assumption will be domain-dependent, and concepts can capture this. Repeated requests in the context of fraud alerts, for example, must be different from those in the context of sales promotions. 

Sometimes whether a design feels dark will vary from user to user. Our framework can capture this by recognizing that users' expected concepts may differ. Consider preselection of a recurring purchase option. On Amazon's platform, for example, such preselection is becoming common for household items that are typically bought repeatedly. Users may reasonably differ on what concept they expect here, with most users (we imagine) seeing such preselection as an insidious attempt by Amazon to sell more, and others (perhaps those who set up repeat purchases often) seeing this as an emerging subscription concept.

\textbf{Limitations:} While our framework accurately describes the taxonomy's dark patterns, a weakness is in describing nuances of the visual design. Other approaches to understanding dark patterns that are more UI-focused provide more specific language to discuss the implications of visual design choices and how those choices impact the user \cite{mathur_dark_2019, di_geronimo_ui_2020}. While our approach's strength lies in describing underlying semantics and functionality, it leans on broader mapping principles for identifying visual issues which may be less precise. For example, our framework can identify that "hard to cancel" violates the usability mapping principle, but it does not point to exact UI elements that cause the  violation.

\newcolumntype{L}{>{\raggedright\arraybackslash}p}
\begin{table*}[t]
    \footnotesize
    {\def\arraystretch{2}
    \begin{tabular}{L{0.1\textwidth} L{0.25\textwidth} L{0.1\textwidth} L{0.175\textwidth} L{0.275\textwidth}}
  \textbf{Dark Pattern} & \textbf{Source Example} & \textbf{Concepts}  & \textbf{Relevant property} & \textbf{Reason for violated expectations} \\ \hline

Comparison Prevention & T-Mobile: bundling prices in different ways                                             & Catalog       & State: price        & Mapping: usability (consistency)  \\
Confirmshaming        & MyMedic: “No, I don’t want to stay alive”                                               & Notification  & Action: opt in/out          & Observed mapping of two options does not match the expectation of symmetry \\
Disguised Ads         & Softpedia: advertisements mimic download buttons                                        & Advertisement  & --           & Mapping: usability (conventions) \\
Fake Scarcity         & Shopify: shows fake low stock messages                                                  & Inventory   & State: stock        & Mapping: transparency                           \\
Fake Social Proof     & Beeketing: shows fake social proof (9 customers just bought this item)                  & Orders      & State: count                   & Mapping: transparency                                        \\
Fake Urgency          & Shopify Hurrify: creates countdown timers that reset upon hitting zero                  & Countdown Timer, Catalog    & State: CountdownTimer.remainingTime, Catalog.salePrice      & Expected synchronization between timer and sale is not implemented                                                  \\
Forced Action         & LinkedIn: required giving access to all email contacts to use site                      & Account, Contact  &  Synchronization: (Account.create, Contact.share)              & Implemented synchronization does not match expectation that consenting to sharing is independent from account creation \\
Hard to Cancel        & New York Times makes it easy to sign up but difficult to cancel                         & Subscription  & Action: subscribe/unsubscribe                & Mapping: usability  \\
Hidden Costs          & Stubhub: advertises low price, goes through many steps, then reveals higher price       & Shopping Cart, Catalog, Order & State: price & Mapping: transparency, usability (consistency) \\
Hidden Subscription   & Figma: when user shares file with editing permissions, a new subscription is charged~   & AccessControl, Subscription  & Synchronization (Subscription.subscribe, AccessControl.share)           & Implemented synchronization does not match expectation that sharing and subscriptions are independent                                     \\
Nagging               & Instagram: aggressively requested users turn on notification, no option to reject fully & Notification  &  Action: enable         & Expected disable action is not implemented                                                 \\
Obstruction           & Facebook: makes it difficult and confusing to reject privacy-invading settings          & PrivacySetting  & Action: accept/reject            & A user who can't navigate the menu observes unexpected lack of reject action       \\
Preselection          & Trump campaign: preselected “make this a monthly recurring donation”~                   & Subscription    & Action: donate, subscribe              & A user who does not notice the box observes unexpected synchronization between donate and subscribe                        \\
Sneaking              & SportsDirect: sneaks item into the user’s shopping cart                                 & Shopping Cart   & Action: add        & Implemented add action is unexpectedly performed by system instead of user            \\
Trick Wording         & Ryanair: used language to make users accidentally book travel insurance                 & Shopping Cart   & Action: add        & A user who does not notice they are adding something observes add action unexpectedly performed by system    \\
Visual Interference   & Tesla: used low contrast text to hide refund policy                                     & Order  & Action: refund                         & A user who does not find the policy observes unexpected lack of refund action
    \end{tabular}
    }
    \caption{Brignull’s Deceptive Designs (April 2023 version) \cite{brignull_deceptive_2023} under our framework. For each deceptive design, we provide an example implementation, list the concepts and their specific state/actions most commonly associated with the design, and explain the way in which the relevant concept component violates user expectations.}
    \label{table:brignulltable}
\end{table*}

\subsection{RQ2: Shopping Cart Case Study}
\label{06-shopping-case-study}

In our second study, we applied our framework to three cases of designs to see if it was capable of articulating distinctions reflected in user experiences.

\textbf{Case 1 — Sports Direct}: Brignull uses Sports Direct, an online sportswear shop, as an example of a clear dark pattern \cite{brignull_dark_2013}. On Sports Direct, after a user adds items to their shopping cart, when they go to checkout they will find that a magazine costing \$1 was added to their cart without their knowledge. To remove it, they must leave the order screen and go back to their shopping cart. This is an unambiguous case of ``sneaking.''

\textbf{Case 2 — Grove Collaborative}
Grove Collaborative (grove.co) is an online household shopping website that includes features that some users find sneaky and problematic but others find them helpful. The design does not appear to have malicious intent; Grove Collaborative seeks to be seen in a positive light, pitching itself as ``a leading socially responsible company'' \cite{noauthor_what_2023}.
One of Grove Collaborative’s services is adding suggested items to the cart for the user. When a user opens the site, there are 13 suggested items already in their cart---some free and some not \cite{noauthor_grove_2023}. When a user checks out, they are automatically enrolled in a recurring subscription service, where suggested items are placed in their cart and automatically checked out each month, described by Grove Collaborative as ``personalized, scheduled Grove shipments, so you’ll never run out of the products you love'' \cite{noauthor_what_2023}.

User reviews highlight the fact that some users find Grove Collaborative’s features to be sneaky, while others find the same features to be helpful and transparent. One disgruntled user lamented that they ``had to always watch [their] cart for extras they sneak in there'' \cite{noauthor_grove_2023}. Another felt that it demanded extra vigilance on their end, writing, ``I don't like the add ons. Sneaky. You have to watch your cart'' \cite{noauthor_grove_2023}. However, other reviews are more positive. One user wrote, ``I actually like the auto-ship. I get an email and app notification a week before my ship date, reminding me to make sure my cart has what I need'' \cite{noauthor_grove_2023}. Users described auto-ship as ``incredibly helpful'' and ``very convenient'' \cite{noauthor_grove_2023}. These varied reviews show that Grove Collaborative is not a clear-cut case of a dark pattern, and that user perceptions vary.

\textbf{Case 3 — StitchFix}: StitchFix is an online clothing company with functionality similar to Grove Collaborative, but that no users seem to perceive as dark. StitchFix puts clothing items in the user's bag based on a style quiz they fill out; when the user checks out, the site charges the user a fee and ships all the items, then the user returns the items they don’t want \cite{noauthor_fashion_2023}. By default, the user is enrolled in a monthly subscription to automatically receive new sets of clothes. The StitchFix mode of shopping bears significant resemblance to the Grove.co mode: suggested items are placed in the user’s cart and shipped each month, providing an automated shopping experience. However, users do not find these autonomous actions sneaky. Negative reviews mention poor customer service, difficulty canceling, issues with payment refunds, and low-quality clothing. However, they do not reflect negative perceptions of the shopping features: amongst about 200 reviews, not one accuses the company of deception in the software design \cite{noauthor_stitch_2023}. 

\subsubsection{Dark Pattern Taxonomies are Insufficient}
\begin{figure}
    \centering
    \includegraphics[width=\linewidth]{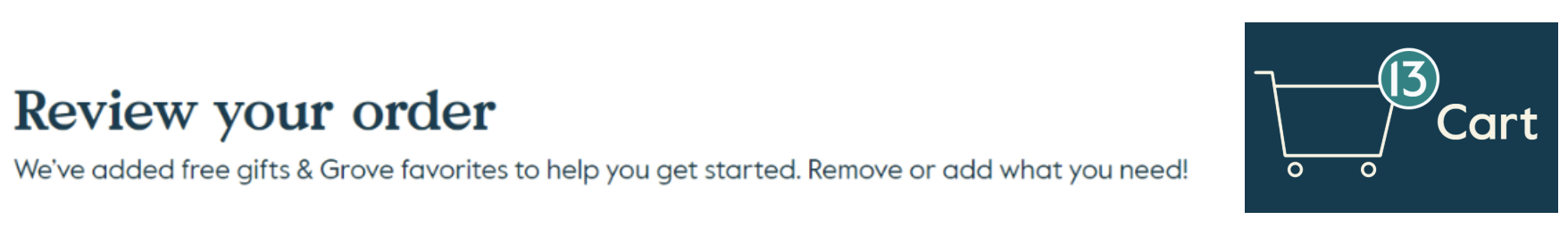}
    \caption{The Grove Collaborative user interface clearly communicates that items have been
added to the cart (through both text and icons) and that they can be removed.
Despite this clear communication, however, some users still feel that the service is
"sneaky".}
    \label{fig:grove}
    \Description{Banner in the Grove checkout user interface. Underneath the title "Review your order", the site explains that they've added free gifts to the cart which can be removed be the user. The cart icon next to the text indicates that 13 free items have been added to the cart.}
\end{figure}

Although these three websites elicited different responses from users, existing taxonomies of dark patterns cannot fully account for these differences. Sports Direct is well described by Brignull's sneaking dark pattern, which is defined as a case where ``the user is drawn into a transaction on false pretenses, because pertinent information is hidden or delayed from being presented to them'' \cite{brignull_deceptive_2023}. But even though quite a few users found Grove Collaborative to be dark, it does not clearly meet this definition. The company markets auto-ship as a valuable part of the service, not something that is snuck behind the user's back. The user interface clearly communicates that items are in the cart and allows the user to remove them, as shown in \autoref{fig:grove}. Before a recurring order is shipped, reminders are sent via text and email both a week before and the day before, giving users the option to cancel or make changes.

But despite this transparency, items being automatically added to the cart by the store feels dark to some users as evidenced by negative reviews. Yet, the existence of positive reviews shows that not all people perceive Grove Collaborative as a dark pattern.  Further, Stitchfix has almost identical functionality to Grove Collaborative, yet does not feel dark to most users. Through the lens of a taxonomy, design techniques either do or do not match a dark pattern definition. A taxonomy does not provide language to discuss why a design may feel subjectively dark to some users but not others.

\subsubsection{Concept Analysis: Evaluating Cases and Generating Solutions}
By using concepts to analyze the ambiguous cases in both procedural and substantive terms, we can understand where important differences arise.

In the Grove Collaborative case, the observed concept does not match the expected concept.
Grove Collaborative's design presents a \texttt{Shopping Cart} concept to the user via conventional shopping cart mappings (shopping cart icon, add to cart buttons, quantity increase / decrease icons, etc.).
Because of this, users expect to be interacting with shopping carts similar to ones they have used in the past.
Most users expect that only the user can initiate the \texttt{add()} action in a shopping cart; the shop calling \texttt{add()} without their involvement breaks this expectation.
In the sneaking dark pattern, the site tries to hide that an item is added to the cart.
Grove Collaborative attempts to avoid this dark pattern by clearly communicating through the interface and email reminders.
Yet, no matter how well the divergence from the norm is communicated, many users will still feel that their expectations are violated.

In contrast, StitchFix avoids breaking users' expectations through the way it communicates its concepts.
When a user opens the site, there is no option to add items to a cart.
Instead, they are directed to the style quiz, which has an interface clearly communicating that stylists will be selecting pieces for the user.
As a result, users do not form an observed concept that includes \texttt{Shopping Cart}, and instead correctly observe that a different concept (which we might call \texttt{PersonalShopper} to connote how it acts on behalf of the user) is in use.
By immediately establishing that they are using a different concept from other shopping sites, StitchFix can implement similar functionality to Grove Collaborative, but without forming expectations they aren't able to fulfill.

We can compare Grove Collaborative and StitchFix both procedurally and substantively to understand how users perceive them differently.
Procedurally, Grove Collaborative causes most users to expect a \texttt{Shopping Cart}, which they implement in a non-standard way.
Users who find Grove Collaborative helpful were likely able to discern that the underlying concept was indeed not a \texttt{Shopping Cart}, and form their expectations accordingly.
StitchFix communicates the \texttt{PersonalShopper} concept, and helps users understand what they should expect it to do.
Users who still interpret the concept as a \texttt{Shopping Cart} may also experience a divergence in observed concept and expected concept.
But from a substantive perspective, this does not necessarily mean that they will perceive this as a dark pattern.
For users who highly value the convenience that the automation brings them, they will feel the divergence is to their benefit---in other words, it is in line with their values.
However, StitchFix users whose observed concept is \texttt{Shopping Cart}, and who do not feel the automation is helpful to them, may perceive StitchFix negatively.

By identifying the parts of an application that violate a standard / expected concept and conflict with users' values and priorities, we can generate fixes. In the case of Grove Collaborative, we identify that the deviation from the standard \texttt{Shopping Cart} comes from the shop having access to \texttt{add()}, and that some users find that this benefits the company at their expense. This can be straightforwardly addressed by only allowing the user to initiate the \texttt{add()} action. However, the shop having access to \texttt{add()} is a fundamental part of Grove's value proposition---Grove seeks to automate and remove the burden of shopping for commonly needed household items. This mismatch between a design goal and the standard concept's functionality is indicative that the concept is not a good fit for the application's needs. To address this, Grove's functionality should be introduced as a distinct concept from \texttt{Shopping Cart}, as StitchFix has done. By establishing a separate concept like \texttt{PersonalShopper}, rather than using standard \texttt{Shopping Cart} UI elements, Grove would no longer be fighting against the expectations users bring from their past \texttt{Shopping Cart} experiences. Alternatively, Grove's suggested additions could be provided by a \texttt{Recommendation} concept, appearing alongside the shopping cart and making it easier for users to add them---but importantly not including them by default in the cart itself.

\subsection{RQ3: Coverage of Popular Websites}
\label{06-popularwebsites}
Through conducting the review of top e-commerce sites, we found that all 20 sites conformed to the core \texttt{Shopping Cart} concept as described by the concept definition and mapping standards in our catalog entry. This finding supports the use of concepts as a lens for studying software design, since many websites re-use identical, well-understood units of functionality. Additionally, our findings support that constructing a concept catalog is feasible, since it is possible to reduce the core functionality of a large corpus of websites to a concise and accurate description. While many iterations were required to \textit{develop} the shopping cart catalog entry, once the \textit{evaluation} of the shopping cart concept began, no changes to the entry were required to describe the core functionality of the examined sites.

In this review, we identified that most websites will implement extensions beyond the minimal functionality in the standard concept. We document these extensions in \autoref{table:extensionstable}. A concept can be extended in an infinite number of ways, and it is impossible to list every possibility in a concept catalog (as discussed in \ref{sec:concept-catalog-entry}). Thus, a concept catalog entry only contains the minimal functionality required for a given concept.
The first part of our dark pattern definition prompts designers to create extensions that do not conflict with the core standard concept. In cases where extensions do conflict, the second part of the definition helps designers differentiate between dark patterns and harmless innovations. 

\definecolor{Gray}{gray}{0.9}
\begin{table}
    \footnotesize
    \begin{tabular}{L{0.2\linewidth} L{0.2\linewidth} L{0.2\linewidth}}
  \textbf{Website} & \textbf{Standard Concept} & \textbf{Extensions} \\ \hline
\rowcolor{Gray}
amazon.com & \checkmark & 1,2,3,4 \\
samsung.com & \checkmark & 2,4,5 \\
\rowcolor{Gray}
ebay.com & \checkmark & 2,4 \\
etsy.com & \checkmark & 1,2 \\
\rowcolor{Gray}
aliexpress.com & \checkmark & 1,3,4,6 \\
walmart.com & \checkmark & 1,2,5,9 \\
\rowcolor{Gray}
apple.com & \checkmark & 5 \\
shein.com & \checkmark & 2,3,8,9 \\
\rowcolor{Gray}
temu.com & \checkmark & 1,3 \\
homedepot.com & \checkmark & 1,2,5,7,9,10 \\
\rowcolor{Gray}
flipkart.com & \checkmark & 2,4 \\
trendyol.com & \checkmark & 3 \\
\rowcolor{Gray}
target.com & \checkmark & 1,2,9 \\
nike.com & \checkmark & - \\
\rowcolor{Gray}
hm.com & \checkmark & - \\
t-mobile.com & \checkmark & 2 \\
\rowcolor{Gray}
lowes.com & \checkmark & 1,2 \\
att.com & \checkmark & 1,2,7,10 \\
\rowcolor{Gray}
zara.com & \checkmark & - \\
wayfair.com & \checkmark & 1,2 \\
\end{tabular}
\caption{Shopping cart concept coverage of popular websites. All of the top 20 most popular e-commerce websites in the US implement our standard shopping cart concept definition, although most websites implement extensions beyond the core functionality (\autoref{table:extensionstable}). According to our dark patterns definition, websites that only contain the standard concept are acceptable. However, any extensions can also be evaluated for darkness using the substantive portion of our definition.}
\label{table:websitetable}
\end{table}

\begin{table}
    \footnotesize
    \begin{tabular}{L{0.03\linewidth} L{0.15\linewidth} L{0.65\linewidth}}
    \textbf{\#} & \textbf{Extension} & \textbf{Description} \\ 
    \hline
1 & Multiple quantity add & \texttt{add()} is augmented with a quantity parameter to allow adding more than one of an item to the cart at a time \\
2 & Save for later & Include state \texttt{saved} to maintain items that the user wants to checkout in a later order. Users can move items freely between \texttt{items} and \texttt{saved}. \\
3 & Partial checkout & Include state \texttt{selectedForCheckout} to maintain items that the user wants to checkout now. Users can select/deselect items in \texttt{items} to be added to/removed from \texttt{selectedForCheckout}, and during \texttt{checkout()} only items in \texttt{selectedForCheckout} are added to the order. \\
4 & Buy now & Include action \texttt{buyNow()} that is synced with \texttt{fulfill()} in order, allowing users to checkout a single item immediately. \\
5 & Estimated total & Include state \texttt{estimatedCheckoutTotal} to maintain the estimated additional shipping fees and taxes that will be added to \texttt{subtotal} during the checkout process. \\
6 & Make account to add to cart & Compose cart with an authorization concept such that users can only access cart actions like \texttt{add()} and \texttt{checkout()} if they have made an account. \\
7 & Share cart & Include state \texttt{viewers}, and compose cart with an authorization concept so that users in \texttt{viewers} can view (but not edit) the contents of \texttt{items}. Only the \texttt{owner} can add other users to \texttt{viewers}. \\
8 & Cost until free shipping bar & The UI displays \texttt{freeShipping - subtotal}. The state \texttt{freeShipping} is from the order concept containing the minimum \texttt{subtotal} the cart must have to qualify for free shipping. \\
9 & Quick add & The UI contains a normal add button on the product information page and a quick add button on the catalog display page where users can add an item without opening its full product information. Both buttons use the \texttt{add()} action under the hood. \\
10 & Empty cart & The UI contains a checkout button and an empty cart button. Both buttons use the \texttt{checkout()} action under the hood. \\
\end{tabular}
\caption{Common extensions to the basic shopping cart concept, found in the list of top 20 e-commerce websites (\autoref{table:websitetable})}
\label{table:extensionstable}
\end{table}

\subsection{Limitations}
\label{sec:study-limitations}
The three studies have some limitations, including:

\textbf{No input / usage from practicing designers: } These RQs focus on evaluating whether the framework and the catalog have the ability to fulfill their respective goals when used by experts who fully understand how to apply them. However, they do not evaluate the approach with practicing designers and therefore cannot capture the benefits and limitations of the framework for practitioners. 

\textbf{No comparison against underlying psychological techniques:} While much of the work in the field of dark patterns has been identifying the underlying psychological techniques employed by dark patterns, in this paper we do not evaluate how those techniques apply to the introduced framework. Therefore, while RQ1 evaluates the applicability of the framework to existing patterns, the framework may not capture nuances that work on underlying techniques highlight.

\textbf{Evaluating website functionality for only a single concept:} We evaluated whether websites contain common shopping cart functionality, but do not extend this evaluation to additional concepts within the sites. Therefore, the possibility remains that not all concepts are so uniformly implemented, which would partially limit the utility of the catalog.

\textbf{Focus on a single type of dark pattern in RQ2:} The shopping case studies focused on evaluating the functionality of adding items to the user's cart, but both StitchFix and Grove Collaborative may have additional dark patterns present in their design that are not evaluated.

\section{Discussion}

\subsection{Merits}
\label{07-merits}

In reflecting on this work, we first highlight the respects in which we believe it contributes to the large body of research on dark patterns and offers some new opportunities.

\textbf{Beyond dark patterns}. Catalogs of dark patterns, along with halls of shame that show instances of these patterns, have been invaluable in shining a light on unethical practices, and play an important role in deterring them. Nevertheless, the volume of research that attempts to build a principled foundation for dark patterns \cite{mathur_what_2021, ahuja_conceptualizations_2022, gray_dark_2018} suggests that catalogs of bad examples are not sufficient. Elucidating the underlying psychological strategies is very helpful in understanding how dark patterns work, but this may not provide an actionable approach. In short, as argued in the introduction, designers need to be able to build confidently on \textit{positive} patterns with the confidence that their work will not be vulnerable to new dark patterns being defined that indict design elements that were previously regarded as acceptable.

\textbf{Defining user expectations}. Following prior work \cite{mathur_what_2021, gray_dark_2018}, we ground our definition of darkness in the deviation between the behavior of the software and the user's expectations. But whereas prior work has taken the existence of these expectations (or, put another way, of a default choice architecture) as a given, our approach recognizes that these expectations are not obtainable by some abstract or psychological analysis but are simply reflections of the way applications currently work. Using concepts we are able to pin down how applications "currently work," and to do so in a modular manner, identifying recurring units of functionality that appear in multiple applications.

\textbf{Procedural and substantive criteria}. Like prior work \cite{mathur_what_2021}, our criterion for darkness couples together the notion of deviation from expectation with an evaluation of whether that deviation actually harms users. In this paper, we have grounded this duality in ethical theories that distinguish procedural and substantive ways of reasoning \cite{zong_bartleby_2022}, and have argued that both approaches are needed.

\textbf{Darkness at two levels}. While dark patterns have focused primarily on  attempts to deceive users in the user interface itself (for example, pushing them to press one button over another, or making a feature hard to find) \cite{brignull_dark_2010, mathur_dark_2019, soe_circumvention_2020, di_geronimo_ui_2020, nouwens_dark_2020}, darkness is to be found also in the underlying semantics of an application. Excessive notifications, for example, that are not generated in response to important events but are rather designed to entice the user into greater engagement, represent a deeper kind of darkness that is not apparent in the design of the user interface per se. Our approach addresses darkness at both levels, using \textit{concepts} as a yardstick for underlying behavior, and using \textit{mapping principles} to indict attempts to mislead in the user interface presentation. 

\textbf{The concept catalog}. Our approach offers a new way to evaluate darkness that we believe can be readily applied by designers as a tool in helping them think about the ethics of design. To be applied more widely and systematically though, we have argued for catalogs of standard concepts that would represent industry consensus on the standard behaviors of recurring units of functionality and standard ways to present them. 

Such catalogs might be produced by industry consortia, or even by individual companies, just as design systems are currently presented as a means of standardizing user interface practices. Individual companies can use concept catalogs both internally and externally. Internally they can be used as guides for design and development work, to help alignment across roles and departments, and to promote consistency across features and products. Externally, they can be used as articulations of the standards the company is committed to following. Indeed, in the area of eCommerce, something like this already exists, with companies like Shopify offering not only APIs but also recommended guidelines for their appropriate use.  Non-profits might play a role in advocating for positive designs by curating catalogs too, just as the W3C consortium published standards for accessibility design. 

Just as we did in creating the Shopping Cart concept catalog entry, organizations wanting to create a concept catalog can start by surveying existing designs, to understand the behavior users are familiar with. This process would not be a purely objective one; it would involve debate over what should be included in the base concept, such as what parts of a common design are problematic and should not be included and whether certain extensions are required or merely preferred. By publishing and debating the details of shared, standard concepts, industry segments can not only promote best practices and make it easier for designers to do socially beneficial work, but might also help indemnify well-meaning designers who, if accused by regulators of dark design, could point to their use of standard concepts in defense.

\subsection{Possible Limitations}
\label{07-limitations}

We now move to addressing some challenges that highlight potential weaknesses in our approach.

\textbf{Concepts may be culture-dependent}. Consider the hypothetical case of two friends out to eat at a restaurant, one American and one European. At the beginning of their meal, they ask the waiter for water for the table. When they receive the check, they find a line item charging them for the water. To the American, this feels outrageous and deceptive, an extra charge being snuck onto the bill so that the restaurant could make more money. But to the European, who is accustomed to paying for water, it is nothing of note. The behavior of the restaurant was the same, but the expectations of the two customers varied, resulting in this feeling like a dark pattern to one but not the other. This example suggests that concepts are not universal, but that they will likely vary across cultures. It also seems likely that sometimes the same generic concept will vary across domains. The \texttt{Notification} concept, for example, although having common behavior in the context of a financial application and a social media application would likely have slight different forms, reflecting differing expectations of privacy and urgency.

\textbf{Users are not uniform in their perception of harm}. As discussed in Section \ref{04-deviation}, evaluating whether a deviation is detrimental to a user's interests involves a substantive analysis that cannot be completely objective and universal, but which must depend on subtle factors relating the users and the context of use. This might mean that in some cases it will be harder to argue whether a particular deviation from a standard concept is acceptable or not. A straightforward resolution of this problem is unlikely, since users will always have different views on what kinds of behavior are acceptable. Most users, for example, are willing to provide their email addresses for authentication and are then enraged if they are automatically signed up for promotional mailings; others are not bothered, so long as unsubscribing is easy. At the very least, it seems that concepts might provide a useful language within which to articulate these differences, and a concept catalog might include annotations that warn that certain variants are unacceptable to some users.

\textbf{Should being standard make a concept less dark?} One argument against our approach is that we essentially give a free pass to design elements that have become standardized, because designs that incorporate them faithfully cannot be accused of deviation from standard practice. On the one hand, this license seems essential, since otherwise it seems that all advertising would have to be regarded as dark. Nevertheless, it is perhaps troubling that our approach would seem not discourage the perpetuation of bad designs. The \texttt{Newsfeed} concept that users have become accustomed to in social media apps such as Facebook essentially grants complete freedom to the developer to choose which posts a user should see. Many users would prefer to have greater control, limiting posts to friends and followers, sorting them chronologically, etc. Arguably a concept such as \texttt{Newsfeed} is inherently dark, as its very design is motivated by goals of engagement and marketing that are not aligned with the interests of users. At the very least, our approach has the merit of highlighting such concepts, and arguing that they should be made explicit in a concept catalog. The process of articulating and debating such concepts might help mitigate the damage that they cause and nudge the field in a better direction.

In tackling this problem, we encourage differentiating between when a design will feel dark to a user (which is tied more closely to the user expectations) vs. when a design is problematic in a broader sense (which is tied to the normative values and expectations of society). To better understand the former, further work is needed to understand how user expectations shift to become accustomed to problematic design. For the latter, we encourage researchers to identify the normative values they are seeking to uphold (as Mathur et al. recommend) \cite{mathur_what_2021}, to define standard concepts that uphold those values, and to identify when the normative values (and therefore, the ideal standard concept) of different stakeholder groups conflict.

For concept catalogs to be helpful, a creation process must be developed that avoids the influence of stakeholders who seek to create standards that have dark patterns built into them. Potential routes for creating concept catalogs with broader impact include catalogs being developed by governmental agencies (such as how NIST creates standards for various industries \cite{noauthor_standards_2016}) or by industry groups (such as how the PCI Security Standards Council “develops and drives adoption of data security standards \cite{noauthor_standards_2023}). 

\textbf{Can a new design be dark?} In our framework, a new design cannot be indicted, however sneaky, if the manipulative elements are couched within concepts that appear to be completely unfamiliar. Unless a design element is tied to a known concept, our approach offers no way to identify what the expectations of the user will be, and therefore cannot be accused of deviation from a norm. In responding to this, one might first note that the existing catalogs of dark patterns reference known design elements and not new ones. Most design work actually involves reuse of existing concepts, and brand new concepts are surprisingly rare. When a new concept is invented, however, there will be a period during which users become acclimatized to that new concept and our approach will have little to say about whether it contains dark elements. Later recognition that a concept has become standardized might reflect a consensus that the concept is ethically acceptable. Perhaps, again, although our approach does indeed seem to fail to fully address this problem, the highlighting of the underlying concept seems to be useful. For example, one might point to Uber's concept of a \texttt{Ride}, which seems to have a dark aspect in the way in which a driver can easily cancel a reservation. This problem seems to have become worse over time, as drivers use the cancellation action more liberally (presumably to abort a booking in favor of another offering higher profit).

\section{Conclusion}\label{07-nextsteps}

We have presented a new approach to dark patterns that complements existing work in several dimensions. Where prior work is mostly focused on the user \textit{interface}, our work starts from the underlying \textit{semantics} of the application; where existing catalogs offer \textit{negative} examples, we have proposed providing \textit{positive} ones; and whereas the emphasis of much work on dark patterns and nudges more generally is on their \textit{mental} qualities, we have focused on describing user's concepts in \textit{behavioral} terms.

We have shown that, as a new lens for understanding dark patterns, a concept-based approach can account for the same kinds of problems that existing approaches identify, albeit via a different route. We have argued that in one key respect our approach fills a crucial missing gap in existing approaches, by providing a notion of expected behavior against which deviations can be judged. 

Whether a concept-based approach can be widely adopted remains to be seen, and some field studies of designers might shed light on this question. We are inspired by the success of design patterns in programming, which share many of the same motivations and qualities as concepts. We are also encouraged by a recent experiment at Palantir \cite{wilczynski_concept-centric_2023} in which concepts were introduced as a new entity type in the company ontology. Since then, over 200 concepts have been identified and documented. Although those concepts may not be sufficiently detailed to act as yardsticks for determining when dark patterns are present, the fact that these concept definitions are now routinely used by many product managers and developers within the company points in a positive direction.

\appendix
\section{appendix}
\label{appendix}
Below we define the guidelines for mapping the shopping cart concept to the user interface, to be used in combination with the mapping principles from Section \ref{mapping_principles}. Interface mapping examples from existing applications can be seen in Figure \ref{mapping_principles}.

\textit{items, price, quantity}
\begin{itemize}
    \item Each item in items should be displayed in the cart UI along with its price and quantity
    \item  The price of each item in items should be the same as its price in the corresponding catalog
    \item  The quantity of each item in items should be less than or equal to the stock available for that item in the corresponding catalog
\end{itemize}

\textit{subtotal}
\begin{itemize}
    \item The subtotal should be displayed in the cart UI.
    \item The conventional meaning of the word “Subtotal” is the total cost of the merchandise before shipping, taxes, and fees. Thus, no other costs in the UI should be labeled “Subtotal” – if additional costs are added, the new totals should be labeled “Total”, “Estimated Total”, “PreTax Order Total”, or a reasonable variation that accurately conveys the fees included in the estimate. 
\end{itemize}
\textit{add()}
\begin{itemize}
    \item There should be an add() button for each item in the corresponding catalog, and the
representation should be consistent.
    \item In addition to managing inventory, the sync between add() and removeFromStock() in catalog prevents users from adding items that are out of stock. An error message should be displayed if a user tries to add() an item that is out of stock, so that the sync is clear to the user. Alternatively, the add() button can be unclickable if the item is out of stock.
\end{itemize}
\textit{changeQuantity()}
\begin{itemize}

    \item There should be a changeQuantity() button for each item in items, and the representation should be consistent
    \item Common instantiations of the UI element representing changeQuantity() are a drop down, a text box with -/+ steppers, or a text box with up/down arrow steppers. The UI should prevent users from choosing a quantity greater than the actual stock available for that item in the catalog. For example, the dropdown could only contain quantities up to the stock available, or the steppers could stop working when the quantity reached is 0 or stock.
    \item Changing the quantity of the item to 0, if allowed, should be synced with the action to remove() the item from the cart, and the UI should reflect this change
\end{itemize}
\textit{remove()}
\begin{itemize}

    \item There should be a remove() button for each item in items, and the representation should be
consistent. 
\end{itemize}
\textit{checkout()}
\begin{itemize}
    \item There should be a checkout() button in the cart UI. 
    \item The checkout experience in a typical e-commerce website is implemented by synchronizing the fulfill() action in order with checkout(). In the cart definition, checkout() is functionally equivalent to emptying the cart, but is named checkout() to emphasize that its primary use is in the checkout experience. However, it is acceptable to include an additional UI element to allow users to empty their cart, which also uses checkout() under the hood to clear items.
\end{itemize}

\begin{figure}[!htb]
    \begin{minipage}[b]{0.37\textwidth}
     \centering
     \includegraphics[width=.9\linewidth]{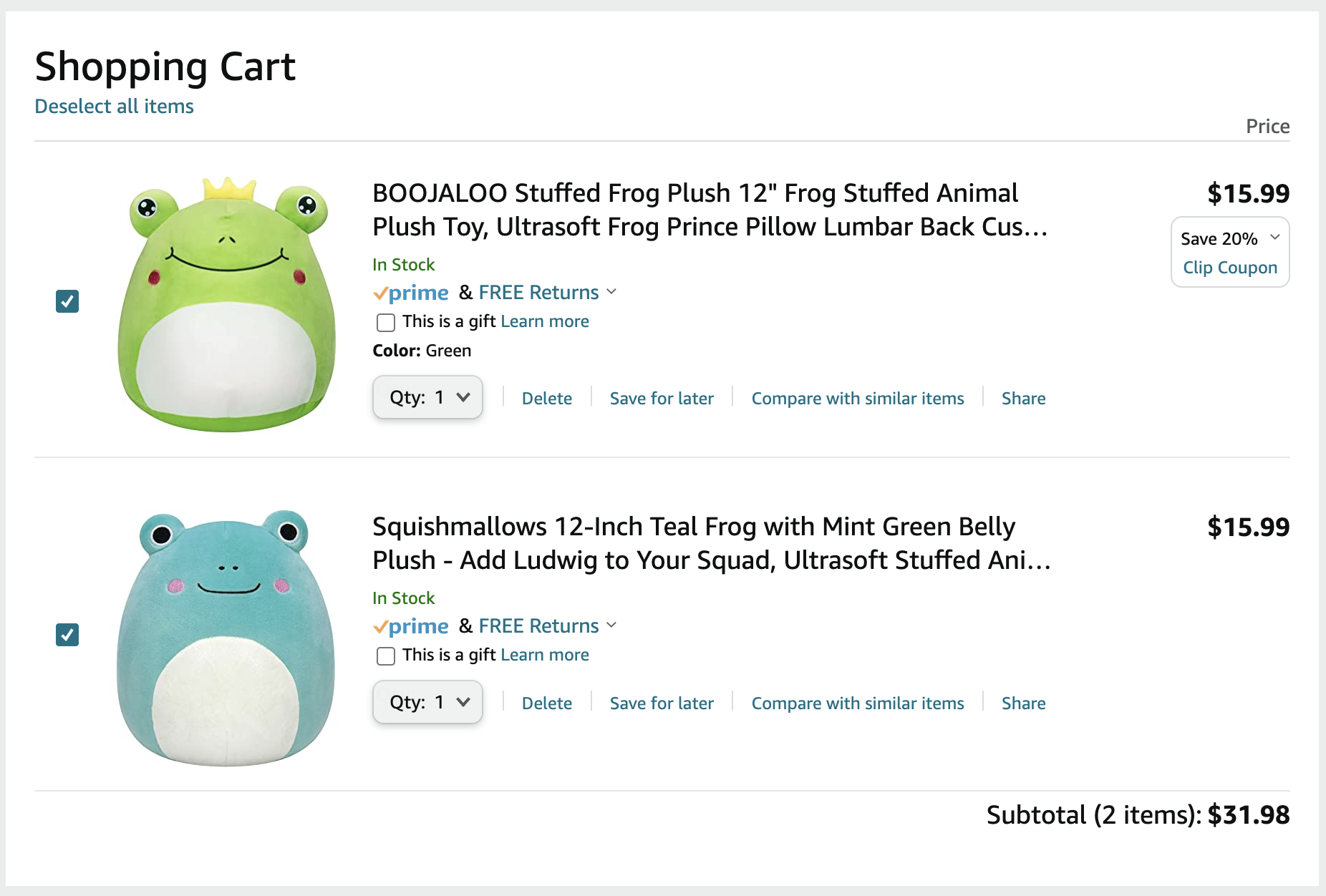} \newline
     (A)
    \end{minipage}\hfill
    \begin{minipage}[b]{0.33\textwidth}
     \centering
     \includegraphics[width=.9\linewidth]{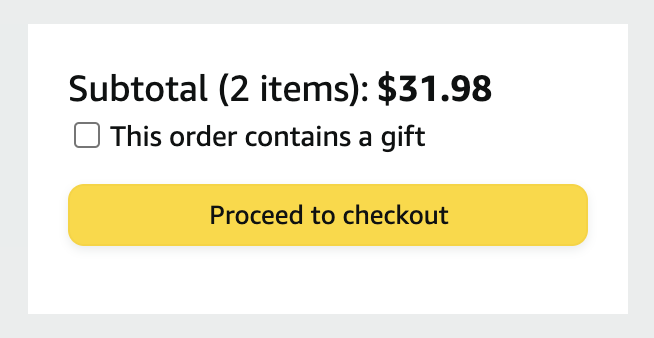} \newline
     (B)
    \end{minipage}
    \begin{minipage}[b]{0.29\textwidth}
     \centering
     \includegraphics[width=.9\linewidth]{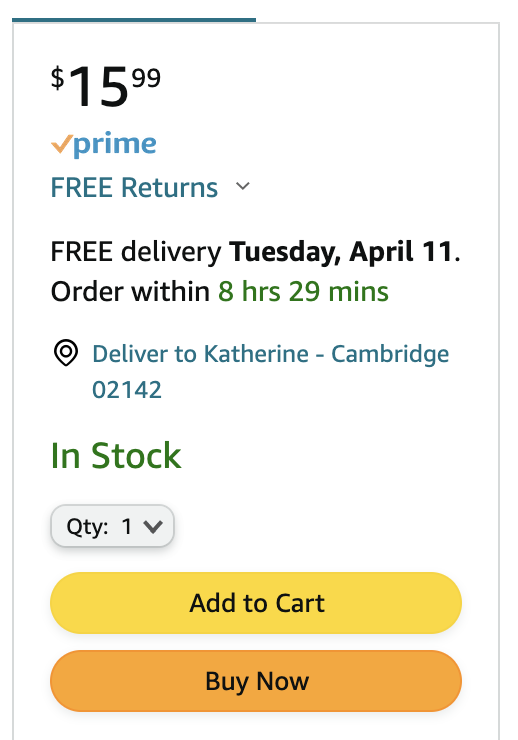} \newline
    (C)
    \end{minipage}
    \caption{Examples of shopping cart user interfaces. (A) items, quantity, price, remove() on amazon.com (B) subtotal, checkout() on amazon.com (C) add() on amazon.com}
    \label{mapping-examples}
    \Description{Screenshots of the Amazon user interface. (A) A sample shopping cart with two items added. (B) Underneath the subtotal is a button to proceed to checkout. (C) Underneath the price of the item are add to cart and buy now buttons.}
\end{figure}

\textbf{Synchronizations}

\textit{Syncs with Catalog}
\begin{itemize}
    \item The p parameter to add() in cart should be price of the item in the catalog
    \item When an item is added to or removed from the cart, the corresponding catalog stock
should be updated. 
\begin{itemize}
    \item add() → removeFromStock()
    \item changeQuantity() → addToStock(), removeFromStock()
    \item  remove() → addToStock()
\end{itemize}
\end{itemize}

\textit{Syncs with Order}
\begin{itemize}
    \item When a cart is checked out, an order should be created
    \begin{itemize}
        \item create() ← checkout()

    \end{itemize}
\end{itemize}

\textit{Syncs with Coupon}
\begin{itemize}
    \item  When a coupon is applied to an item in the cart, the price of that item should be updated
    \begin{itemize}
        \item changePrice() ← apply()

    \end{itemize}
\end{itemize}

\begin{acks}
This work was funded in part by a grant from the CCF division of the National Science Foundation, under grant number 2131541 in the program \textit{Designing Accountable Software Systems} (NSF 21-554). We are grateful also to the other members of the LAChS research team---Jason Goldberg, Naomi Kirimi, Ilaria Liccardi, Geoffrey Litt and Danny Weitzner---who contributed ideas, suggested examples, and were helpful skeptics as we refined our approach.
\end{acks}

\bibliographystyle{ACM-Reference-Format}
\bibliography{concept-catalog}

\end{document}